# Synthetic parity-time symmetry breaking in a single microcavity

**Fangxing Zhang[1], Yaming Feng[2], Li Ge[3*], and Wenjie Wan[1,2*]**

[1]The State Key Laboratory of Advanced Optical Communication Systems and Networks, University of Michigan-Shanghai Jiao Tong University Joint Institute,
Shanghai Jiao Tong University, Shanghai 200240, China

[2]MOE Key Laboratory for Laser Plasmas and Collaborative Innovation Center of IFSA,
Department of Physics and Astronomy, Shanghai Jiao Tong University, Shanghai 200240, China

[3]Department of Physics and Astronomy, College of Staten Island, the City University of New York, NY 10314,
and the Graduate Center, CUNY, New York, NY 10016, USA

*Corresponding authors: Wenjie Wan wenjie.wan@sjtu.edu.cn or Li Ge li.ge@csi.cuny.edu

**Non-Hermitian systems based on parity–time (PT) symmetry reveal rich physics beyond the Hermitian regime. So far, realizations of PT-symmetric systems have been limited to the spatial domain. Here we theoretically and experimentally demonstrate PT symmetry in a synthetic spectral dimension induced by nonlinear Brillouin scattering in a single optical microcavity, where electromagnetically induced transparency or absorption in two spectral resonances provides the optical gain and loss to observe a phase transition between two symmetry regimes. This scheme provides a new paradigm towards the investigation of non-Hermitian physics in a synthetic photonic dimension for all-optical signal processing and quantum information science.**

The operator description of quantum mechanics, formulated by Dirac and von Neumann, postulated that all physical observables are real-valued and associated with a Hermitian operator, including the energy and its operator, the Hamiltonian. However, Bender and colleagues theoretically demonstrated in 1998 [1] that a new class of non-Hermitian Hamiltonians can possess a real energy spectrum below a certain phase-transition point, immediately opening up an opportunity for exploring interesting consequences of non-Hermitian physics. These systems are characterized by parity-time (PT) symmetry, with a complex potential $V(x)$ subject to the spatial-symmetry constraint $V(x) = V^*(-x)$. They undergo a phase transition near an "exceptional point", which signals the spontaneous breaking of PT symmetry and turns the real energy spectrum into a complex one. Ten years later, Christodoulides and collaborators proposed and experimentally realized such PT symmetry in optics [2], by utilizing the analogy between the Schrödinger equation in quantum mechanics and the paraxial optical diffraction equation. Complex PT-symmetric potentials were achieved by spatially modulating the optical refractive index with properly balanced gain and loss spatial profiles.

For example, a pair of coupled waveguides, with identical real part and opposite imaginary parts of the refractive index, can display periodic but asymmetric power oscillations between them in the PT-symmetric phase [3]; they can also lead to loss-enhanced transmission in the PT-broken phase, even in a passive waveguide pair without gain [2]. Later, this idea was generalized to show laser self-termination with increasing gain [4,5] and loss-induced lasing [6], and there has been a plethora of other discoveries in non-Hermitian photonics based on PT symmetry [7], including anisotropic transmission resonances [8] and unidirectional invisibility in a single PT-symmetric waveguide [9,10], robust wireless power transfer [11], coexistence of coherent perfect absorption and lasing [12,13,14], single-mode micro-ring laser with predetermined orbital angular momentum [15], and enhanced photonic sensing near an exceptional point [16,17].

So far, these investigations of PT symmetry have been limited to the spatial arrangement of effective gain and loss, which originates from the definition of PT symmetry in its original form, where the parity operation is a mirror reflection. While there are theoretical work extending the definition of the parity operator in the context of non-Hermitian Hamiltonians [18], i.e., to embrace any linear operator including rotation and inversion [19], such generalized PT symmetry and its exceptional point have not been observed experimentally. The importance of utilizing such a "synthetic parity operator" lies in a broader scope of non-Hermitian physics based on PT symmetry, which offers a more versatile approach beyond spatially balanced gain and loss. Indeed, a related concept proposed by Fan and coworkers [20], i.e., synthetic dimension in photonics, has opened up a new territory for investigating optical effective gauge fields [21] and topological photonics [22].

This additional, synthetic dimension is enabled by coupling modes with different frequencies or angular momenta, achieved by imposing direct temporal modulation in resonant structures such as a whispering-gallery-mode (WGM) optical microcavity [23]. Besides this "external" modulation approach, it is also well known that light can be converted into other frequencies in these microcavities through resonance-enhanced nonlinearities, including stimulated Raman scattering [24], Brillouin scattering [25], four-wave mixing [26] and optomechanical oscillations [27]. This "internal" modulation approach offers a new route to realize a synthetic frequency dimension, which has led to the observation of some striking photonic dynamics, e.g., optically induced transparency and slow light [28,29]. To realize PT symmetric photonics with the synthetic dimension in the microcavity remains a big challenge, and yet be demonstrated up to date. It is highly desired to eliminate the spatial parity like the prior works with two-coupled microcavities [30,31] and synthesize an additional frequency domain through nonlinear conversion, for a new platform to investigate such non-Hermitian physics.

In this work, we theoretically and experimentally demonstrate PT symmetry breaking in this synthetic spectral dimension in one single optical microcavity. Unlike previous explorations of the original PT symmetry, here two *spectrally* separated optical resonances in a *single* microcavity are coupled by nonlinear frequency conversion through stimulated Brillouin scattering (SBS), without the need of having a second microcavity in the spatial PT case. These two nonlinear-coupled resonances can interfere with each other and result in either electromagnetically induced transparency (EIT) or absorption (EIA), similar to their counterparts in atomic physics. By using a detuning technique developed specifically for this framework, we are able to observe a striking phase transition between the broken and unbroken phases of the synthetic PT symmetry at an exceptional point. These results expand the scope of non-Hermitian optics into the synthetic spectral dimension, providing a new paradigm to exploit the benefits of PT symmetry and the emerging pseudo-Hermitian physics.

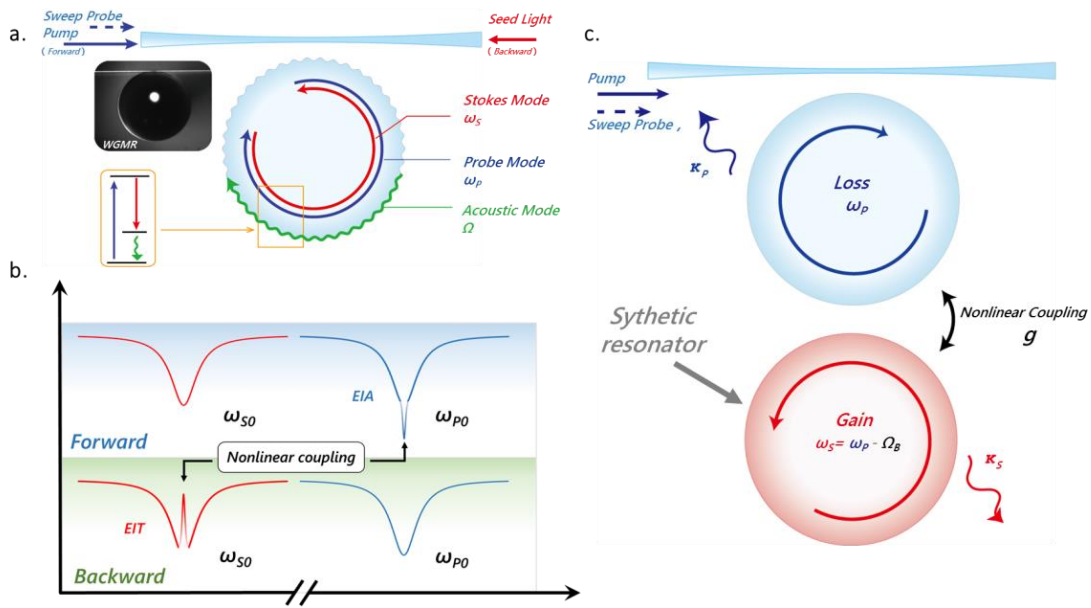

**Figure 1:** Conceptual illustration of a synthetic PT-symmetric optical microcavity. (a) Nonlinear Brillion scattering in a WGM microcavity with two optical modes, i.e., the probe and Stokes modes, together with an acoustic mode. Lower inset: Energy transition diagram. Upper inset: SEM of one high-Q microsphere microcavity coupled to the tapered fiber. (b) Nonlinear gain and loss processes manifested as EIT and EIA in the transmission spectra. The EIT peak and the EIA dip need to be measured separated using a probe in the backward and forward direction, respectively. (c) Schematics showing the synthetic PT symmetry along the frequency dimension. $\kappa_{p,s}$ are the linear losses of the probe and Stokes modes.

Our proposed spectral PT symmetric system is shown schematically in Fig. 1. Here an incident probe wave at frequency $\omega_p$ propagates from left to right (forward) in a tapered fiber and evanescently tunnels into the neighboring microsphere resonator. By sweeping its frequency at low power, clockwise (CW) WGM resonances can be excited sequentially. The resulting dips in the transmission spectrum (Fig. 1b, top; without the sharp EIA dip on the right) are given by inverted Lorentzians well separated from each other, indicating no nonlinear coupling between them. However, if the power of the probe increases above certain threshold, it can stimulate the generation of Stokes photons at a lower frequency $\omega_s$ and coherent acoustical phonons at frequency $\Omega = \omega_p - \omega_s$ through an electrostrictive-induced SBS process [14]. This process has a low threshold if $\omega_{p,s}$ and $\Omega_B$ are close to their respective optical and acoustic WGM resonances [25]. In our experiment the Stokes photons are confined in a counterclockwise (CCW) WGM mode (Fig. 1a) to fulfill the conservation law of momentum. The induced Brillouin dynamical grating (BDG) [32] from the generated acoustical wave now serves as a channel to nonlinearly couple the two optical resonances involved, providing effective gain to the Stokes photons and loss to the probe photons, based on which the synthetic PT symmetry is realized (Fig. 1c).

To keep a stable nonlinear coupling between the probe and Stokes photons while sweeping the probe frequency, we in fact use a strong pair of the pump at $\omega_1$ close to $\omega_p$ (Fig. 2a). and the seed at $\omega_2$ near $\omega_s$ to lock the BDG, which allows us to control the frequency of the generated acoustic phonons (details to be elucidated later). Here to discuss the synthetic PT symmetry of the probe and Stokes waves, the dynamics of the strong pump and seed can be eliminated [see Supplemental Material (SM)], leading to a coupled-mode theory for the amplitudes of the probe wave ($A_p$) and Stokes wave ($A_s$) in the microsphere:

$$i\frac{\partial A_p}{\partial t} = (-i\kappa_p - \Delta_p)A_p - g\, A_s$$
$$i\frac{\partial A_s}{\partial t} = (-i\kappa_s - \Delta_s)A_s + g^*A_p \qquad (1)$$

Here $\kappa_{p,s}$ are their respective cavity decay rates due to internal absorption, scattering, radiation loss and external coupling. $\Delta_{p,s} = \omega_{p,s} - \omega_{p0,s0}$ are the frequency detunings from their neighboring WGM frequencies (denoted by $\omega_{p0,s0}$), which are related by $\Delta_p - \Delta_s = \Omega - (\omega_{p0} - \omega_{s0})$. $\Delta_p$ is controlled directly by scanning the probe frequency, while $\Delta_s$ is tuned indirectly by changing the seed frequency via the SBS generated phonons at frequency $\Omega = \omega_1 - \omega_2$. The most crucial coupling term $g$ can also be controlled by the strength of the BDG (see SM).

As we mentioned before, the standard PT symmetry in optics is satisfied by coupling two or more optical modes in spatially separated resonators or waveguides with different effective gain and loss. Using the simplest case without losing generality, the non-Hermitian Hamiltonian can be written as a 2x2 matrix in the basis of two uncoupled modes [3]:

$$H_{PT} = \begin{bmatrix} i\gamma & \kappa \\ \kappa & -i\gamma \end{bmatrix} \quad (2)$$

Here the gain and loss parameter $\gamma$ and the linear coupling $\kappa$ are both real, and the parity operator $P$ is a mirror reflection given by the first Pauli matrix $\sigma_x$, which exchanges the spatial positions of these two optical modes. The PT symmetry is evident through the relation $[H_{PT}, PT] = 0$, where the time reversal operator $T$ is given by the complex conjugation.

To compare Eq. (1) to this standard PT Hamilton, we rewrite it in the basis $A_p = a_p\, e^{i(\Delta + i\kappa_p)t}, A_s = a_s\, e^{i(\Delta + i\kappa_s)t}$:

$$i\frac{\partial}{\partial t}\begin{pmatrix} a_p \\ a_s \end{pmatrix} = \begin{pmatrix} -\delta & -g \\ g^* & \delta \end{pmatrix}\begin{pmatrix} a_p \\ a_s \end{pmatrix} \equiv H_{SPT}\begin{pmatrix} a_p \\ a_s \end{pmatrix} \quad (3)$$

where $\Delta \equiv (\Delta_p + \Delta_s)/2$ is the average detuning and $\delta = (\Delta_p - \Delta_s)/2$. Unlike $H_{PT}$, there is no explicit gain or loss in this non-Hermitian Hamiltonian; the effective loss to the probe photons and effective gain to the Stokes wave are represented inexplicitly by the off-diagonal nonlinear coupling terms. Therefore, it is clear that $H_{SPT}$ does not commute with the original $PT$ operator, with $P = \sigma_x$ now exchanges the spectral position of the probe wave and its Stokes wave in the same cavity. However, it can be shown that $H_{SPT}$ still satisfies a generalized PT symmetry $[H_{SPT}, P_sT] = 0$, with a synthetic parity operator defined by

$$P_s = \frac{ig_r\sigma_y - \delta\sigma_z}{\sqrt{\delta^2 - g_r^2}}. \quad (4)$$

Here $\sigma_{y,z}$ are the other two Pauli matrices, $g_r = Re[g]$, and the denominator is introduced to satisfy $(P_sT)^2 = 1$. The two eigenvalues of $H_{SPT}$ are given by:

$$\lambda_\pm = \pm\sqrt{\delta^2 - |g|^2}. \quad (5)$$

Obviously, when $|\Delta_p - \Delta_s| = 2|\delta| > 2|g|$ (referred as the unbroken PT phase), these two eigenvalues are real and different, resulting in two spectrally separated resonances. It is important to note that they are not the two optical WGM modes in the rotating frame. Instead, they indicate that the probe and the Stokes wave inside the cavity now hybridize to form two dynamical eigenmodes, one with mixed frequencies $\Omega_{p,s} + \sqrt{\delta^2 - |g|^2}$ and the other $\Omega_{p,s} - \sqrt{\delta^2 - |g|^2}$.

The phenomenon is independent of our pump-seed scheme and would occur if we excite the acoustic phonons mechanically to induce the BDG. In the other regime of the PT broken phase, i.e., $|\Delta_p - \Delta_s| < 2|g|$, these two eigenfrequencies are complex conjugates, with the same real part (spectral center) but different imaginary parts (decay rates). $|\Delta_p - \Delta_s| = 2|g|$ gives the exceptional point, where the two dynamical eigenmodes (instead of the two optical WGMs) become identical, i.e., the probe (Stokes) wave oscillates at the frequency $\Omega_p$ ($\Omega_s$). This is an important difference between exceptional points in linear and nonlinear optical systems [30,31]. We should mention that these spectral properties also indicate that $H_{SPT}$ is pseudo-Hermitian [33], satisfying $\eta^{-1} H_{SPT} \eta = H_{SPT}^\dagger$ with $\eta = \sigma_z$. What's more, $H_{SPT}$ anti-commutes with $\sigma_x T$ and hence also has anti-PT symmetry [34-36]. This symmetry is the reason that $\lambda_\pm$ are purely imaginary in the PT broken phase [37].

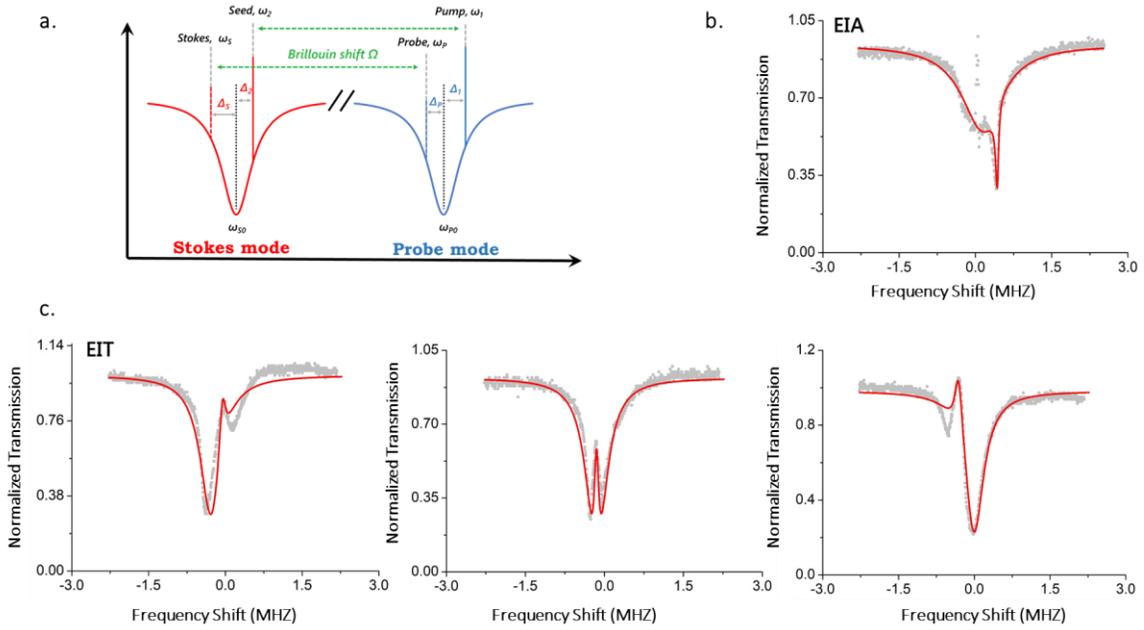

**Figure 2:** Electromagnetically-induced transparency and absorption in a synthetic PT-symmetric microcavity. (a) Illustration of the pump-seed set up in the experimental, where a strong pump beam at $\omega_1$ and a seed at $\omega_2$ are launched to induce a Brillouin dynamic grating. A weak probe in the forward direction is used to measure the EIA near the pump (b) and synthetic PT breaking (Fig. 3). In a separate experiment a backward probe is used instead to scan near the seed frequency (c), which demonstrates the tuning of the phonon frequency via the pump-seed controlled BDG.

Experimentally, we first investigate the nonlinear coupling induced gain and loss near the probe and its Stokes resonance. Under the steady-state condition, the transmission spectra of the probe in the forward direction in the fiber and that of the Stokes resonance in the backward direction can be derived from Eq. (1) to show occurrences of EIT and EIA in Fig. 2 (details in SM). They arise from the interference between the probe and Stokes resonances through nonlinear coupling. Take the EIT for example (Fig. 2c). The narrow gain peak induced near the Stokes resonance (in the backward direction) is enhanced by the optical resonance at the pump frequency through a BS process, which has a relative board spectral width ~30MHz (inversely proportional to acoustic lifetime) [32] and wider than the optical resonance ~1MHz in our experiment. EIT or EIA under such a broad gain bandwidth was considered unlikely, but we note that nonlinear coupling in fact narrows the gain linewidth resonantly to be on the order of the optical resonance, similar to a FWM enabled optically induced transparency we have demonstrated previously [26]. Similar spectral profiles can be found in the probe's EIA transmission in the forward direction near the pump resonance as shown in Fig. 2b, where the induced absorption dip also has similar linewidth as its optical resonance. These results validate the nonlinear coupling between the probe and its Stokes resonance, offering us a synthetic platform to realize PT symmetry.

To observe synthetic PT symmetry breaking, we vary the two detunings $\Delta_{p,s}$ systematically. This approach is very different from the investigations of the standard PT symmetry in twin-coupled microcavities [30,31], where the detuning between the two optical resonances, one in each microcavity, needs to be avoided by all means to achieve PT symmetry. Because both the pump-seed pair and the probe-Stokes pair are related through the SBS generated phonons, i.e., $\omega_1 - \omega_2 = \omega_p - \omega_s = \Omega$ (see Fig. 2a), the critical parameter $|\Delta_p - \Delta_s|$ for synthetic PT symmetry breaking is equal to $|\Omega - (\omega_{p0} - \omega_{s0})|$. Therefore, this critical parameter is tuned only by the seed frequency in our pump-seed scheme with a fixed pump frequency and does not actually depend on the probe frequency. We use the latter property to our advantage: for each seed frequency, we can then map out the central frequencies and linewidths of the two dynamical modes predicted by Eq. (4) by simply scanning close to the pump frequency.

In Fig. 2c-e, we show the tuning of the induced transparency window near the Stokes mode (i.e., $\Delta_s$) by detuning the seed beam in our pump-seed scheme, which is both effective and systematic. Note that when $|\Delta_s|$ is large, this peak evolves into an asymmetric Fano lineshape (Fig. 2c) similar to prior works [26]. In the meantime, we do not observe any visible shift of the EIA spectrum in Fig. 2b, because it highly depends on the strong pump's detuning fixed during the whole measurement.

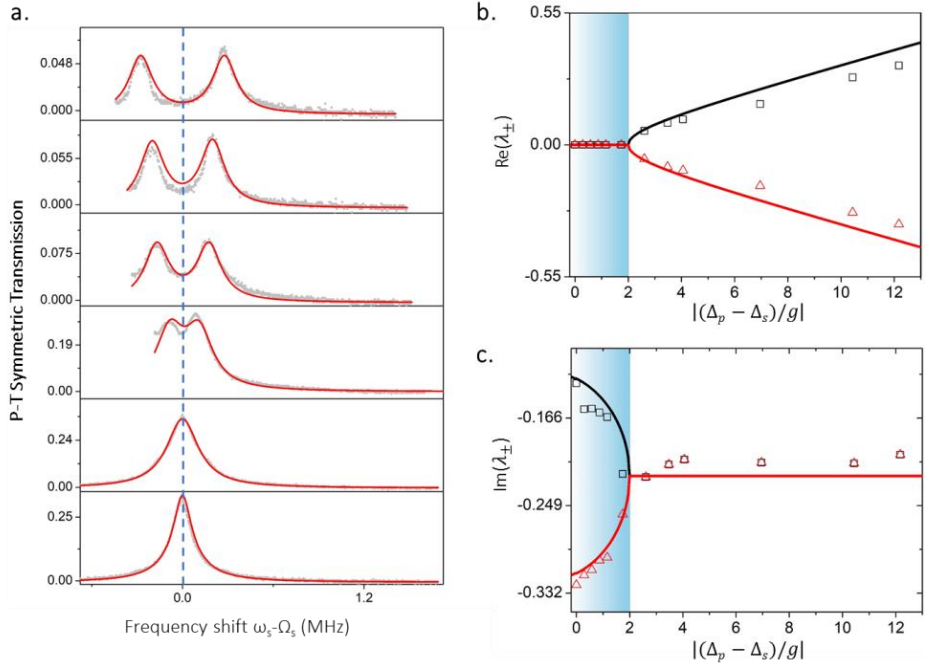

**Figure 3:** Experimental observation of synthetic PT symmetry breaking. (a) Spectral evolution of the probe transmission in the forward direction by tuning the phonon frequency via the seed. $\Omega_s$ is the un-split frequency defined as $\Omega_s = (\omega_{P0} + \omega_{S0} - \Omega)/2$ [SM]. The pump and seed lasers are fixed at 15.22mW and 69.4µW, respectively, while the power of the weak probe is 3.48 µW. The real parts (b) and imaginary parts (c) of the dynamical eigenfrequencies measured in the synthetic PT system. The exceptional point at $|(\Delta_p - \Delta_s)/g| = 2$ marks the phase transition boundary. The solid lines are obtained from the theoretical model in Eq. (5).

The breaking of synthetic PT symmetry in our system is observed using a lock-in amplification technique [SM] on the probe beam, which can detect the presence of any dynamical mode predicted by Eq. (5). Meanwhile, we manage to precisely tune the seed beam in order to control $|\Delta_p - \Delta_s|$ and keep a relative constant $|g|$. During this tuning process, both $\Delta_{p,s}$ and the effective gain $|g|$ are calibrated in SM. Initially, the two modes represented by $\lambda_\pm$ in Eq. (5) are split by about 0.6 MHz under the condition $|(\Delta_p - \Delta_s)/g| \approx 12$ As the ratio becomes smaller in the unbroken PT phase, so does the frequency difference of the two dynamical modes (Fig. 3b), which agrees well with the theoretical prediction given by Eq. (5). Meanwhile, the linewidths of both modes maintain a constant value (Fig. 3c), also agreeing with Eq. (5). Physically, the effective nonlinear gain is not strong enough to compete with the cavity loss such that two modes are relative independent. However, as the detuning factor gradually reduces to $|\Delta_p - \Delta_s| \approx 2|g|$, i.e., the exceptional point, the two split modes begin to merge into one single mode, where the frequency splitting vanishes (Fig. 3b). Above the EP, the two modes remain aligned but differ in their linewidths, which can be extracted from Fig. 3a using the superposition of two Lorentizan lineshapes. The measured result again verifies the prediction of Eq. (5), which completes our demonstration of a synthetic PT symmetry breaking in a single microcavity.

In conclusion, we have realized PT symmetry in a synthetic spectral dimension and observed its spontaneous breaking using nonlinear Brillouin scattering in a single optical microcavity. Our work opens a new frontier at the junction of two emerging fields in physics, namely synthetic dimension and non-Hermitian physics based on PT symmetry. Our results also offer a unique perspective on the relationship between non-Hermitian physics and nonlinear optical processes, which can be extended to investigate a wide range of phenomena including second-harmonic generation, parametric amplification [38] and four-wave mixing [39].

**Acknowledgements:**
This work was supported by National key research and development program (Grant No. 2016YFA0302500, 2017YFA0303700); National Science Foundation of China (Grant No. 11674228, No. 11304201, No. 61475100); Shanghai MEC Scientific Innovation Program (Grant No. E00075).

**Reference**
1. Boettcher, S. & Bender, C. M. Real spectra in non-Hermitian Hamiltonians having PT symmetry. *Phys. Rev. Lett.* **80**, 5243_5246 (1998).
2. Guo, A. et al. Observation of PT-symmetry breaking in complex optical potentials. *Phys. Rev. Lett.* **103**, 093902 (2009).
3. Rüter, Christian E., Makris, K. G., El-Ganainy, R., Christodoulides, D. N., Segev, M., & Kip, D. Observation of parity–time symmetry in optics. *Nature Phys.*, **6**, 192-195 (2010).
4. Liertzer, M., Ge, L., Cerjan, A., Stone, A., Türeci, H., & Rotter, S. Pump-induced exceptional points in lasers. *Phys. Rev. Lett.*, **108**, 173901-0. (2012).
5. El-Ganainy, R., Khajavikhan, M., & Ge, L. Exceptional points and lasing self-termination in photonic molecules. *Phys. Rev. A*, **90**, 013802 (2014).
6. Peng, B., Ozdemir, Ş. K., Rotter, S., Yilmaz, H., Liertzer, M., & Monifi, F., et al. Loss-induced suppression and revival of lasing. *Science*, **346**, 328-332 (2014).
7. Feng, L., El-Ganainy, R., & Ge, L. Non-hermitian photonics based on parity–time symmetry. *Nat. Photonics,* **11**, 752-762 (2017).
8. Ge, L., Chong, Y. D., & Stone, A. D. Conservation relations and anisotropic transmission resonances in one-dimensional PT-symmetric photonic heterostructures. *Phys. Rev. A*, **85**, 023802 (2012).
9. Feng, L. et al. Experimental demonstration of a unidirectional reflectionless parity time metamaterial at optical frequencies. Nature Mater. 12,108_113 (2012).

10. Zin, L., Ramezani, H., Kottos, T., Eichelkraut, T., Christodoulides, D. N., & Hui, C. Unidirectional Invisibility Induced by PT-Symmetric Periodic Structures. *Phys. Rev. Lett.* **106**, 213901 (2011).
11. Assawaworrarit, S., Yu, X., & Fan, S. Robust wireless power transfer using a nonlinear parity–time-symmetric circuit. *Nature*, **546**, 387-390 (2017).
12. Longhi, S. Pt-symmetric laser-absorber. *Phys. Rev. A*, **82**, 9583-9588 (2010).
13. Wan, W., Chong, Y., Ge, L., Noh, H., Stone, A. D., & Cao, H. Time-reversed lasing and interferometric control of absorption. *Science* **331**(6019), 889-892 (2011).
14. Wong, Z. J., Xu, Y. L., Kim, J., O"Brien, K., Wang, Y., & Feng, L., et al. Lasing and anti-lasing in a single cavity. *Nat. Photonics*, **10**, 796–801 (2016).
15. Miao, P., Zhang, Z., Sun, J., Walasik, W., Longhi, S., & Litchinitser, N. M., et al. Orbital angular momentum microlaser. *Science*, **353**, 464-467 (2016).
16. Wiersig, & Jan. Enhancing the sensitivity of frequency and energy splitting detection by using exceptional points: application to microcavity sensors for single-particle detection. *Phys. Rev. Lett.*, **112**, 203901 (2014).
17. Chen, W., Şahin Kaya Özdemir, Zhao, G., Wiersig, J., & Yang, L. Exceptional points enhance sensing in an optical microcavity. *Nature*, **548**, 192-196 (2017).
18. Deng, J. W., Guenther, U., & Wang, Q. H. General PT-symmetric matrices. arXiv preprint arXiv:1212.1861 (2012).
19. Bender, C. M., Meisinger, P. N., & Wang, Q. All Hermitian Hamiltonians have parity. *J. Phys. A: Math. Gen.* **36**, 1029 (2003).
20. Yuan, L., Lin, Q., Xiao, M., & Fan, S. Synthetic dimension in photonics. *Optica.* **5**, 1396 (2018).
21. Yuan, L., Shi, Y., & Fan, S. Photonic gauge potential in a system with a synthetic frequency dimension. *Opt. Lett.* **41**, 741 (2016).
22. Lin, Q., Xiao, M., Yuan, L., & Fan, S. Photonic Weyl point in a two-dimensional resonator lattice with a synthetic frequency dimension. *Nat. Commun.* **7**, 13731 (2016).
23. Vahala, K. J. Optical microcavities. *Nature* **424**, 839–846 (2003).
24. Spillane, S. M., Kippenberg, T. J., & Vahala, K. J. Ultralow-threshold raman laser using a spherical dielectric microcavity. *Nature* **415**, 621-623 (2002).
25. Bahl, G., Zehnpfennig, J., Tomes, M., & Carmon, T. Stimulated optomechanical excitation of surface acoustic waves in a microdevice. *Nat. Commun.* **2**, 403. (2011).
26. Zheng, Y., Yang, J., Shen, Z., Cao, J., Chen, X., & Liang, X., et al. Optically induced transparency in a micro-cavity. *Light-Sci. Appl.* **5**, e16072 (2016).
27. Kippenberg, T. J., & Vahala, K. J. Cavity opto-mechanics. *Opt. Express* **15**(25), 17172-205 (2007).
28. Kim, J., Kuzyk, M. C., Han, K., Wang, H., & Bahl, G. Non-reciprocal Brillouin scattering induced transparency. *Nat. Phys.*, **11**, 275 (2015).
29. Dong, C. H., Shen, Z., Zou, C. L., Zhang, Y. L., Fu, W., & Guo, G. C. Brillouin-scattering-induced transparency and non-reciprocal light storage. *Nat. Commun.* **6**, 6193 (2015).

30. Peng, B., Özdemir, Ş. K., Lei, F., Monifi, F., Gianfreda, M., Long, G. L., Fan, S., Nori, F., Bender, C. M., & Yang, L. Parity–time-symmetric whispering-gallery microcavities. *Nat. Phys.* **10**, 394 (2014).
31. Chang, L., Jiang, X., Hua, S., Yang, C., Wen, J., Jiang, L., Li, G., Wang, G., & Xiao, M. Parity–time symmetry and variable optical isolation in active–passive-coupled microresonators. *Nat. Photonics* **8**, 524 (2014).
32. Feng, Y., Zhang, F., Zheng, Y., Chen, L., Shen, D., Liu, W., & Wan, W. Coherent control of acoustic phonons by seeded Brillouin scattering in polarization-maintaining fibers. *Opt. Lett.* **44**, 2270-2273 (2019).
33. Mostafazadeh, A. Pseudo-Hermiticity versus PT symmetry: the necessary condition for the reality of the spectrum of a non-Hermitian Hamiltonian. *J. Math. Phys.* **43**, 205-214 (2002).
34. Ge, L., & Türeci, H. E. Antisymmetric PT-photonic structures with balanced positive-and negative-index materials. *Phys. Rev. A*, **88**, 053810 (2013).
35. Youngsun, C., Choloong, H., Woong, Y. J. & Ho, S. S. Observation of an anti-pt-symmetric exceptional point and energy-difference conserving dynamics in electrical circuit resonators. *Nature Commun*, **9**, 2182- (2018).
36. Peng, P., Cao, W., Shen, C., Qu, W., Wen, J., & Jiang, L., et al. Anti-parity–time symmetry with flying atoms. *Nature Phys,* **12**, 1139 (2016).
37. Ge, Li. Symmetry-protected zero-mode laser with a tunable spatial profile. *Phys. Rev. A*, **95**, 023812 (2017).
38. El-Ganainy, R., Dadap, J. I., & Osgood, R. M. Optical parametric amplification via non-Hermitian phase matching. *Opt. Lett.*, **40**, 5086-5089 (2015).
39. Ge, L., & Wan, W. Pseudo-Hermitian Transition in Degenerate Nonlinear Four-Wave Mixing. arXiv preprint arXiv:1603.05624 (2016).